\RequirePackage[hyphens]{url}
\documentclass[
reprint,
groupedaddress,
amsmath,amssymb,
aps,
]{revtex4-1}

\usepackage{graphicx}
\usepackage{siunitx}
\usepackage{dcolumn}
\usepackage{bm}
\usepackage{kantlipsum}
\PassOptionsToPackage{hyphens}{url}\usepackage{hyperref}
\usepackage[mathlines]{lineno}
\usepackage{cleveref}
\usepackage{physics}

\begin{document}

\preprint{APS/123-QED}

\title{A Compact, Mobile, Low-Threshold Squeezed Light Source}
\author{J.\,Arnbak}
\author{C.\,S.\,Jacobsen}
\author{R.\,B.\,Andrade}
\author{X.\,Guo}
\author{J.\,S.\,Neergaard-Nielsen}
\author{U.\,L.\,Andersen}
\author{T.\,Gehring}
\email{tobias.gehring@fysik.dtu.dk}
\affiliation{Center for Macroscopic Quantum States bigQ, Department of Physics, Technical University of Denmark}

\date{\today}

\begin{abstract}
Strongly squeezed light finds many important applications within the fields of quantum metrology, quantum communication and quantum computation. However, due to the bulkiness and complexity of most squeezed light sources of today, they are still not a standard tool in quantum optics labs. We have taken the first steps in realizing a compact, high-performance \SI{1550}{\nano\meter} squeezing source based on commercially available fiber components combined with a free-space double-resonant parametric down-conversion source. The whole setup, including single-pass second-harmonic generation in a waveguide, fits on a small breadboard and produces \SI{9.3}{\decibel} of squeezing at a \SI{5}{\mega\hertz} sideband-frequency. The setup is currently limited by phase noise, but further optimization and development should allow for a 19" sized turn-key squeezing source capable of delivering more than \SI{10}{\decibel} of squeezing.   

\end{abstract}

\maketitle

\section{\label{sec:intro}Introduction}

Squeezed quantum states of light are a ubiquitous resource in numerous applications associated with quantum sensing, quantum communication and quantum computation~\cite{Andersen2016,Andersen2015,Lvovsky2015,Weedbrook2012,Pirandola2018}. One of the most celebrated examples is the application of squeezed light to improve the sensitivity of gravitational wave interferometers, thereby extending the volume in space within which gravitational events can be observed.~\cite{Schnabel2010,LSC2011,Aasi2013}. A recent impressive improvement in observable volume is the eight-fold increase by the detection of 6 dB squeezed light in the gravitational wave detector GEO600~\cite{Geo6002019}. Quantum-enhanced sensitivity can also be achieved with squeezed light in tracking the motion and estimating bio-physical parameters of single living cells~\cite{Taylor2016,Taylor2013,Taylor2014}.

Apart from quantum sensing, squeezed light also has applications in quantum cryptography to extend the secure communication distance~\cite{Madsen2012}, to improve the cryptographic security~\cite{Gehring2015} and to enable the implementation of quantum secure basic cryptographic primitives~\cite{Furrer2018}. Finally, squeezed light has recently been found to be a viable resource for photonics continuous variable quantum computing due to development of new quantum error correcting codes~\cite{Gottesman2001, Menicucci2014} and due to the inherent scalability of the squeezed light source~\cite{Weedbrook2012,Marshall2016, Yokoyama2013,Larsen2019, Chen2014}.

All of the above mentioned applications would naturally benefit from a compact, mobile and robust squeezed light source producing an appreciable amount of squeezing. However, in most experiments to date there have been a sharp trade-off between achieving high degrees of  squeezing and the compactness (and thus robustness and transportability) of the source.

On one hand, quantum states have been significantly squeezed by up to \SI{15}{\decibel} (that is, a reduction of 97\% of the vacuum noise) using a nonlinear crystal embedded in an optical cavity~\cite{Vahlbruch2016,Takeno2007,Eberle2010,Mehmet2011,Shi2018,Schonbeck2018}, but the associated experimental setups have very large footprints and are not easily transportable due to the need for multiple mode-cleaning cavities for the pump and local oscillator to maximize the squeezing and cavity based second harmonic generation to supply the pump which easily exceeds \SI{100}{\milli\watt}. Due to this immobility, the squeezed light sources are often built up around the application.

On the other hand, compact and mobile squeezed light sources have been constructed using either an optical waveguide in a single-pass configuration~\cite{Mondain2019,Takanashi2019} or as a cavity~\cite{Stefszky2017}, using a micron-sized Silicon Nitride ring cavity~\cite{Dutt2015, Vaidya2019} or using a whispering gallery mode resonator~\cite{Otterpohl2019}, but in these systems the squeezing degree have been limited to maximum \SI{3}{\decibel}. Using the optical Kerr effect in fibers, the production of 2.4\,dB pulsed squeezed light was demonstrated on a mobile platform of $0.3\, \text{m}^2$~\cite{Peuntinger2014}. In all of these latter experiments, compactness has been traded with the squeezing degree. 

The trade-off has been partially settled in a couple of realizations: The free-space squeezed light source -- an optical parametric oscillator -- used in the GEO600 gravitational wave detector has a footprint of \SI{1.1 x 1.3}{\meter}, a weight of approximately \SI{70}{\kilo\gram}, and produces about \SI{10}{\decibel} squeezing~\cite{Vahlbruch2010}.
Finally, a \SI{50 x 60}{\centi\meter} free-space setup producing \SI{6}{\decibel} of two-mode squeezed light has recently been demonstrated~\cite{Wang2015}.
\section{\label{set:setup}Setup}
\begin{figure*}
    \centering
    \includegraphics[width = 0.9\linewidth]{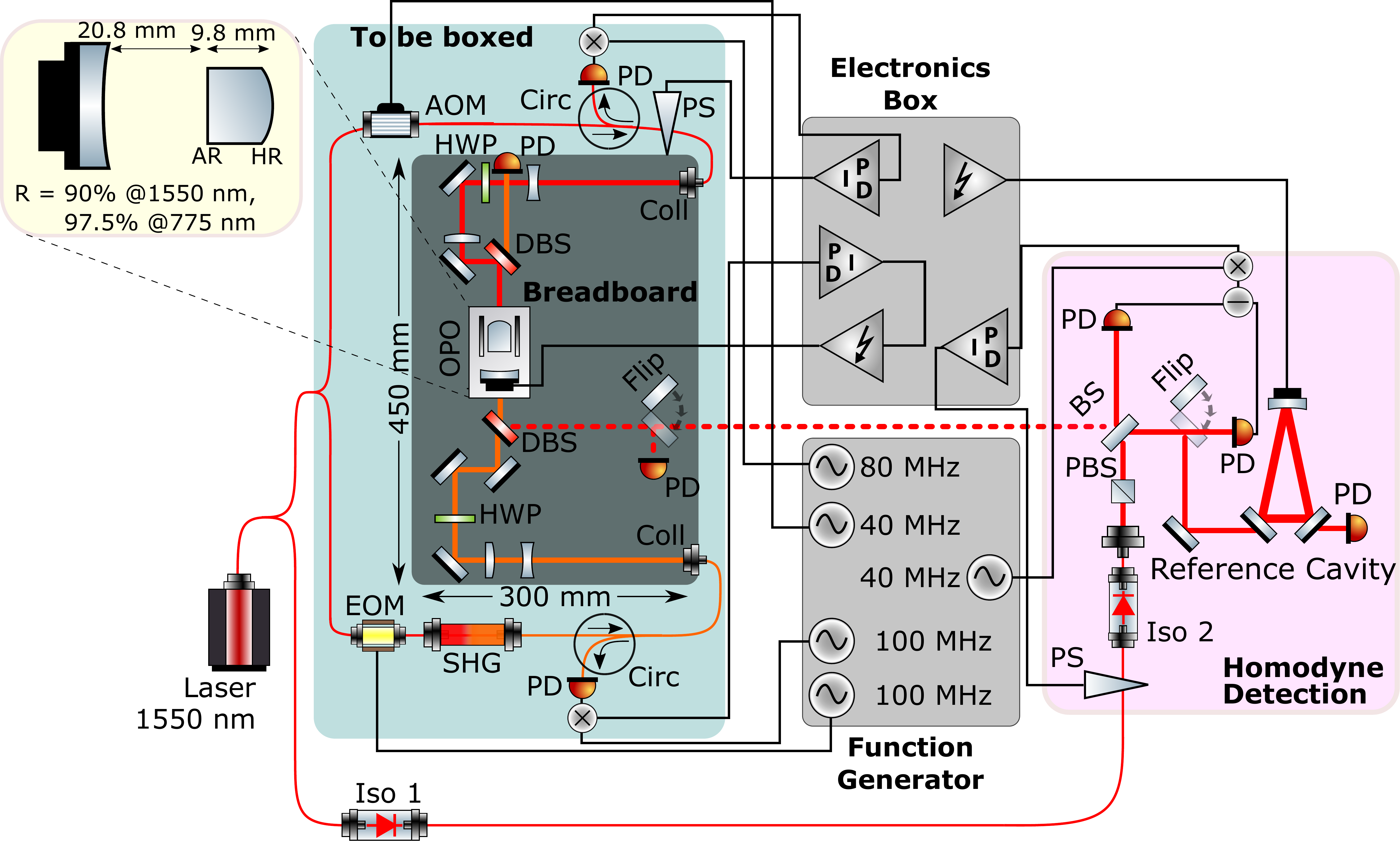}
    \caption{
    Schematic representation of the experimental setup. The free-space part of the setup is placed on a breadboard measuring \SI{30 x 45}{\centi\meter} with the fiber components being placed around for convenience. The plan is to fit both the free-space part and the fiber components (marked by the teal rectangle) into a 19" box.\\
    The \SI{775}{\nano\meter} light, used to generate the squeezed light, is generated in a single-pass LiNbO$_3$ waveguide second-harmonic generator (SHG). A Pound-Drever-Hall lock in the pump path stabilizes the cavity. A coherent-locking scheme for locking the relative phase between pump and local oscillator utilizes a \SI{40}{\mega\hertz} up-shifted pilot tone transmitted together with the squeezed light. The squeezed light is characterized by a balanced homodyne setup that utilizes a reference cavity to help mode-match the squeezed light and local oscillator. Polarization sensitive fiber isolators are inserted along the local oscillator fiber path in order to minimize polarization noise build-up along the fiber path.\\
    AOM: acousto-optic modulator, EOM: electro-optic modulator, BS: (50/50) beam-splitter, DBS: dichroic beam-splitter, OPO: double-resonant optical parametric oscillator, PD: photo-detector, HWP: half-wave plate, PS: phase-shifter, Iso: isolator, Circ: circulator, Coll: collimator, Flip: flip mirror}
    \label{fig:setup}
\end{figure*}

\begin{figure}
    \centering
    \includegraphics[width=8.5cm]{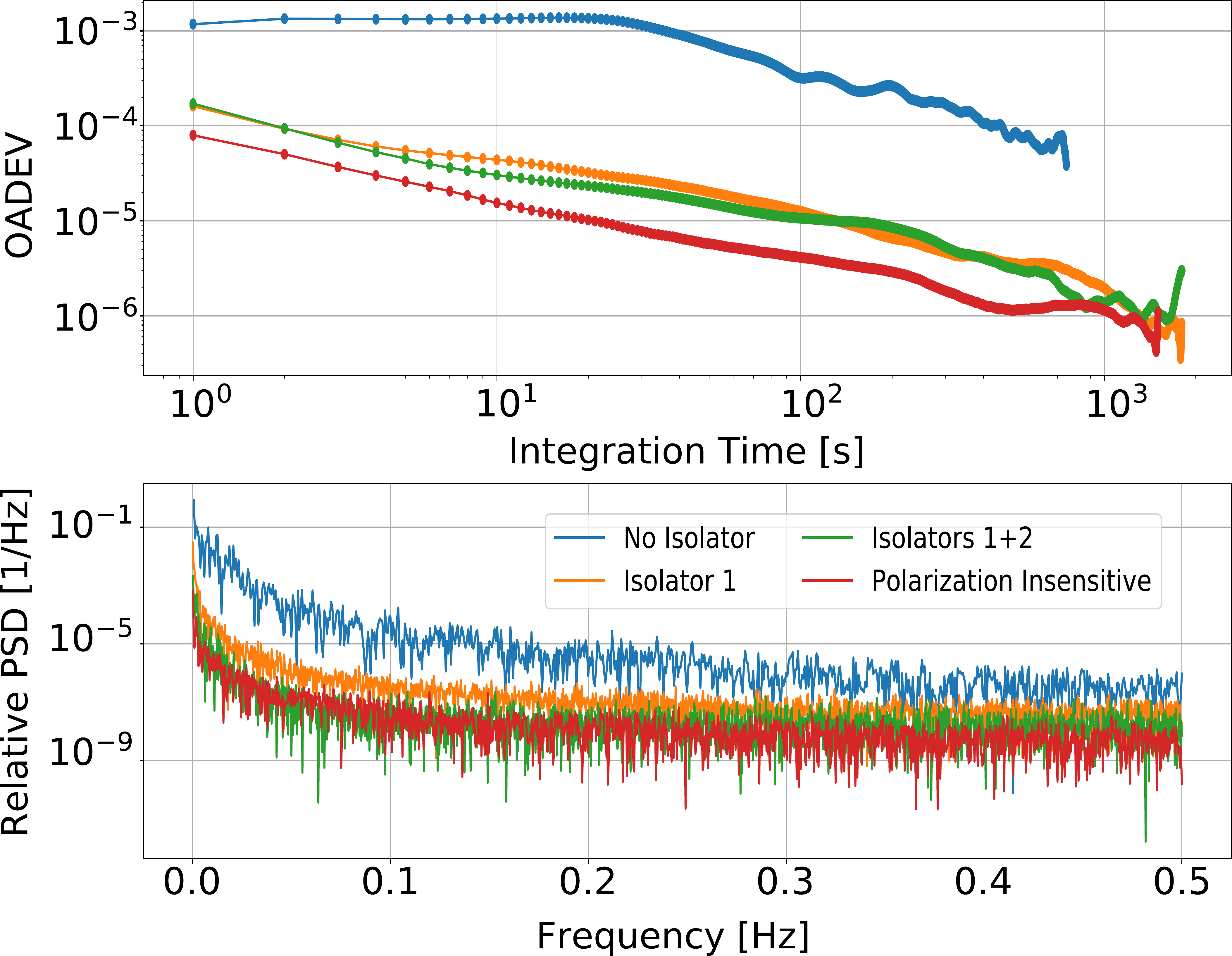}
    \caption{Overlapped Allan deviation (OADEV) and power-spectral density of polarization noise measurements in the local oscillator path. The data was taken at a \SI{1}{\hertz} sampling rate and is normalized to the mean and has mean subtracted. The blue trace is without any polarizing components. The yellow trace is with one polarization sensitive isolator and the green trace is with two isolators. The red trace is polarization insensitive amplitude noise.}
    \label{fig:PSD}
\end{figure}

In this article, we present the construction of a compact squeezed light source with a footprint of \SI{30 x 45}{\centi\meter} producing 9.3\,dB squeezing. While the source itself is a traditional double resonant parametric down-conversion source, we reduced its size by replacing bulky free-space optical components with less effective, but more compact commercially available fiber-alternatives and operate without the use of filter cavities. Only the couplings of the pump beam and the coherent control beam to the nonlinear cavity are obtained by free-space optics, as is the squeezed light output. In particular, the second-harmonic generator consists of a small single-pass waveguide module which produces enough light to saturate our source's low pump power threshold of \SI{5.2}{\milli\watt}---a record for this type of source~\cite{Schonbeck2018}. The setup is complemented with a 19"-sized laser and 19"-sized control electronics which have roughly the same footprint as our source.

Our experiment is presented in Fig.\,\ref{fig:setup}. The \SI{1550}{\nano\meter} squeezed light is generated via parametric down-conversion by pumping a double resonant optical parametric oscillator (OPO) with light at \SI{775}{\nano\meter} below threshold. The OPO itself, as sketched in the yellow inset, is a hemilithic cavity with a planar-convex periodically poled potassium titanyl phosphate (PPKTP) crystal as the non-linear medium. The crystal has a size of \SI{9.8 x 2 x 1}{\milli\meter}, the planar facet is coated with an anti-reflection (AR) coating (R$<$\SI{0.1}{\percent} @\SI{1550}{\nano\meter}, R$<$\SI{0.3}{\percent} @\SI{775}{\nano\meter}), and the curved facet (radius of curvature (ROC) $=$ \SI{10}{\milli\meter}) is coated with a high-reflection (HR) coating (R$=$\SI{99.95}{\percent} @\SI{1550}{\nano\meter}, R$>$\SI{99.5}{\percent} @\SI{775}{\nano\meter}). The output mirror (ROC$=$\SI{25}{\milli\meter}) is coated with a partially reflective coating (R$=$\SI{90}{\percent} @\SI{1550}{\nano\meter}, R$=$\SI{97.5}{\percent} @\SI{775}{\nano\meter}). The cavity has a round-trip optical length of around \SI{77}{\milli\meter} resulting in a free spectral range of approximately \SI{3.9}{\giga\hertz}, a finesse of around 58 for the \SI{1550}{\nano\meter} mode (full-width-half-maximum (FWHM) $\approx$ \SI{66}{\MHz})  and around 200 for the \SI{775}{\nano\meter} mode (FWHM $\approx$ \SI{17}{\MHz}). The double-resonance is achieved by tuning the temperature of the crystal and the length of the cavity with a piezo-electric transducer.

An NKT Photonics E15 BOOSTIK \SI{1550}{\nano\meter} fiber laser supplies the light to the setup. The light is divided into two paths; one serving as the local oscillator in the homodyne measurement setup. The other path is further split into a pumping path and a path for the coherent control beam~\cite{Vahlbruch2016_Coherent}, in the following called the pilot beam.

The pumping path starts with a \SI{1550}{\nano\meter} electro-optic modulator (EOM) to modulate the phase at \SI{100}{MHz} for the OPO Pound-Drever-Hall (PDH) lock. The EOM has a maximum input power of \SI{300}{\milli\watt} and an insertion loss of \SI{3}{\decibel}. This limits the input power into the LiNbO$_3$ second-harmonic generator (SHG) wave-guide module (NTT Electronics WH-0775-000-F-B-C) to \SI{150}{\milli\watt}, resulting in around \SI{8}{\milli\watt} of \SI{775}{\nano\meter} light to be used to pump the OPO.

The pilot path implements a coherent-locking scheme for locking the relative phase between the pump light and the local oscillator by using a fiber acousto-optic modulator (AOM) to up-shift the frequency of a \SI{1550}{\nano\meter} beam by 40 MHz. This pilot field enters the OPO through the HR side and interacts with the pump field through difference-frequency generation. The reflected light is detected and down-mixed with an \SI{80}{\mega\hertz} tone to generate an error signal for locking the phase of the pilot field to the pump field. The locked pilot field is transmitted with the squeezing and beats with the local oscillator. After detection in the homodyne detector, it is down-mixed with a \SI{40}{\mega\hertz} tone to provide an error-signal for locking the phase between the local oscillator and the squeezed field.

Both the pump and the pilot beams are coupled out of the fibers and collimated by fiber collimators. Using two mode-matching lenses and two steering mirrors, each of the free-space beams are coupled into the OPO. All free-space optics are placed on the \SI{30 x 45}{\centi\meter} breadboard which leaves enough space to place all fiber components on it as well. In the actual experiment, the fiber components were not attached to the breadboard out of convenience, but will be placed in the box in the final version of the device.

The squeezed light is characterized in a balanced homodyne setup which is placed on a neighboring breadboard and uses a bright (\SI{10}{\milli\watt}) \SI{1550}{\nano\meter} beam as a local oscillator. The overlap between the squeezed light and the local oscillator is optimized by coupling both fields into a triangular reference cavity. This allows us to achieve a fringe visibility of around \SI{99}{\percent}. The homodyne detector uses InGaAs photodiodes ($\eta_{QE}>$ \SI{99}{\percent}). The photo-electric signal is analyzed with a spectrum analyzer.

\begin{figure}
    \centering
    \includegraphics[width=8.5cm]{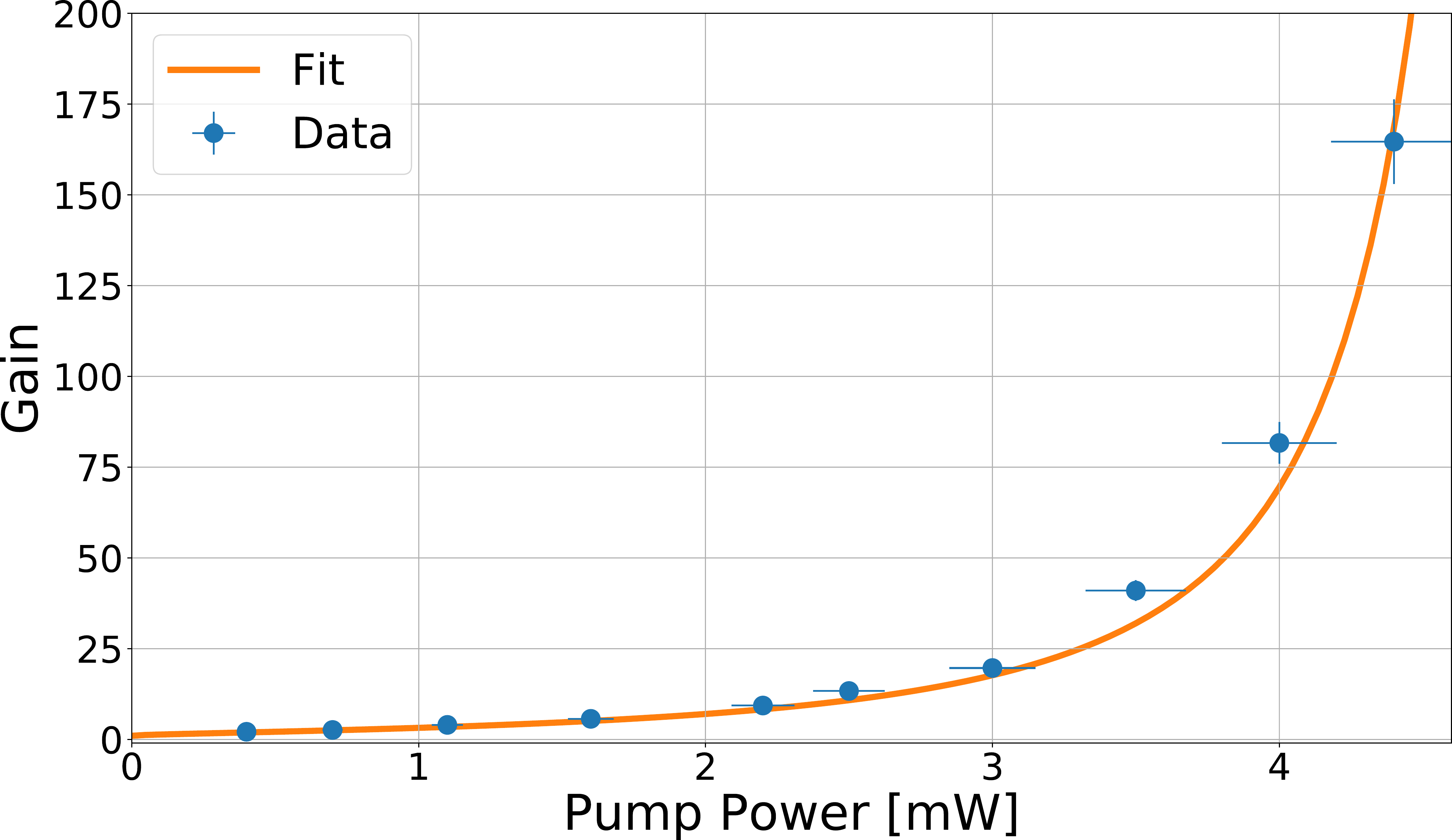}
    \caption{Graph showing experimentally obtained gain values (orange dots) as a function of input pump power. The blue line is a fit to \cref{eq: gain}, and a threshold value of \SI[separate-uncertainty,multi-part-units = single]{5.12(3)}{\milli\watt} is extracted from the model. The error-bars assume a 5\% error on the power.}
    \label{fig:gain}
\end{figure}

\section{Results}
The setup utilizes polarization maintaining (PM) fiber components, and every time two PM fibers are combined in a mating sleeve, a small polarization mismatch can appear due to an imperfect slow-axes alignment. With many components, such a mismatch can build up and result in quite severe polarization noise. This problem can be circumvented by using a polarization stabilizing feed-back loop, but a simpler solution is to insert polarization filtering components along the path, thus preventing polarization mismatch to build up. The downside to this simple solution is the introduction of some excess amplitude noise.

In our setup, the EOM, AOM, SHG and circulators are all polarization filtering components thus preventing mismatch to build up in the pump and pilot paths. For the local oscillator, however, polarization noise  was a problem, and fiber isolators were therefore inserted into the fiber path.

We characterize the effect of inserting polarizing components by inserting a power-meter after a polarizing beam-splitter (PBS) measuring at a sampling rate of \SI{1}{\hertz}. Fig.~\ref{fig:PSD} shows a plot of the overlapped Allan deviation and the power spectral density of these measurements. We measured without any isolators, with isolator 1 and with two isolators, isolators 1 and 2. For their positions see the experimental schematic in Fig.~\ref{fig:setup}. As a baseline, we measure the polarization insensitive amplitude noise. The data is normalized to the mean, and the mean is subtracted in order to get the fluctuations relative to the mean.

From the Allan deviation, we see a large increase in long-term stability of the power when isolator 1 is present. This can also be seen in the power spectral density, with a difference of almost two orders of magnitude between having no polarizing components and having isolator 1. Between having one and two isolators, the difference is negligible for both graphs which is to be expected as no additional fiber components are present after isolator 2. By comparing the green/orange traces with the red trace, it seems that the amplitude noise with isolators is in general a bit higher than the intrinsic noise from the laser. This could be due to small polarization mismatches being converted to amplitude noise by the isolators.

The performance of our setup is first characterized by estimating the pump threshold power via a classical gain measurement of a \SI{1550}{\nano\meter} field interacting with the pump field in the crystal. For this measurement, the fiber AOM is removed and the transmitted power of the \SI{1550}{\nano\meter} field is measured while varying the input power of the \SI{775}{\nano\meter} pump field. The gain is then estimated by comparing with the transmitted \SI{1550}{\nano\meter} light when the pump is blocked. The calculated gain values are plotted in Fig.~\ref{fig:gain}.
\noindent The gain is modelled as~\cite{Aoki2006}  
\begin{align}
    g = \frac{1}{\left(1-\sqrt{\frac{P_{p}}{P_{p}^\text{thr}}}\right)^2},
    \label{eq: gain}
\end{align}
where $P_{p}$ is the input pump power and $P_{p}^\text{thr}$ is the OPO threshold power. From the fit of the experimental gain values, we can extract an OPO threshold power of \SI[separate-uncertainty,multi-part-units = single]{5.12(3)}{\milli\watt}. This shows that even though the system only has \SI{8}{\milli\watt} of pump power available due to lossy fiber components, the double resonance allows the squeezer to utilize its full potential, since squeezing is best generated below threshold.

\begin{figure*}
    \centering
    \includegraphics[width=8.5cm]{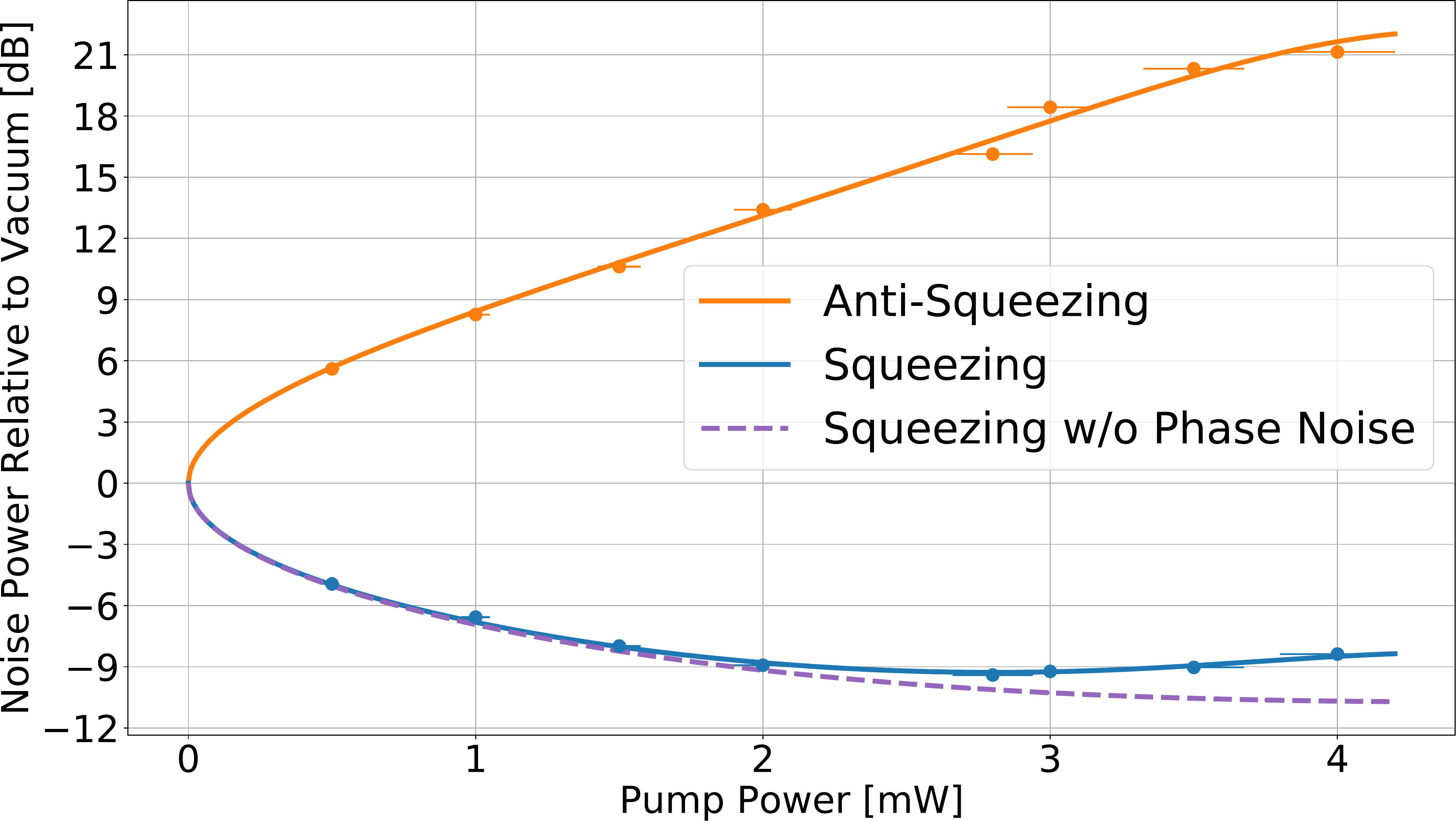}
    \includegraphics[width=8.5cm]{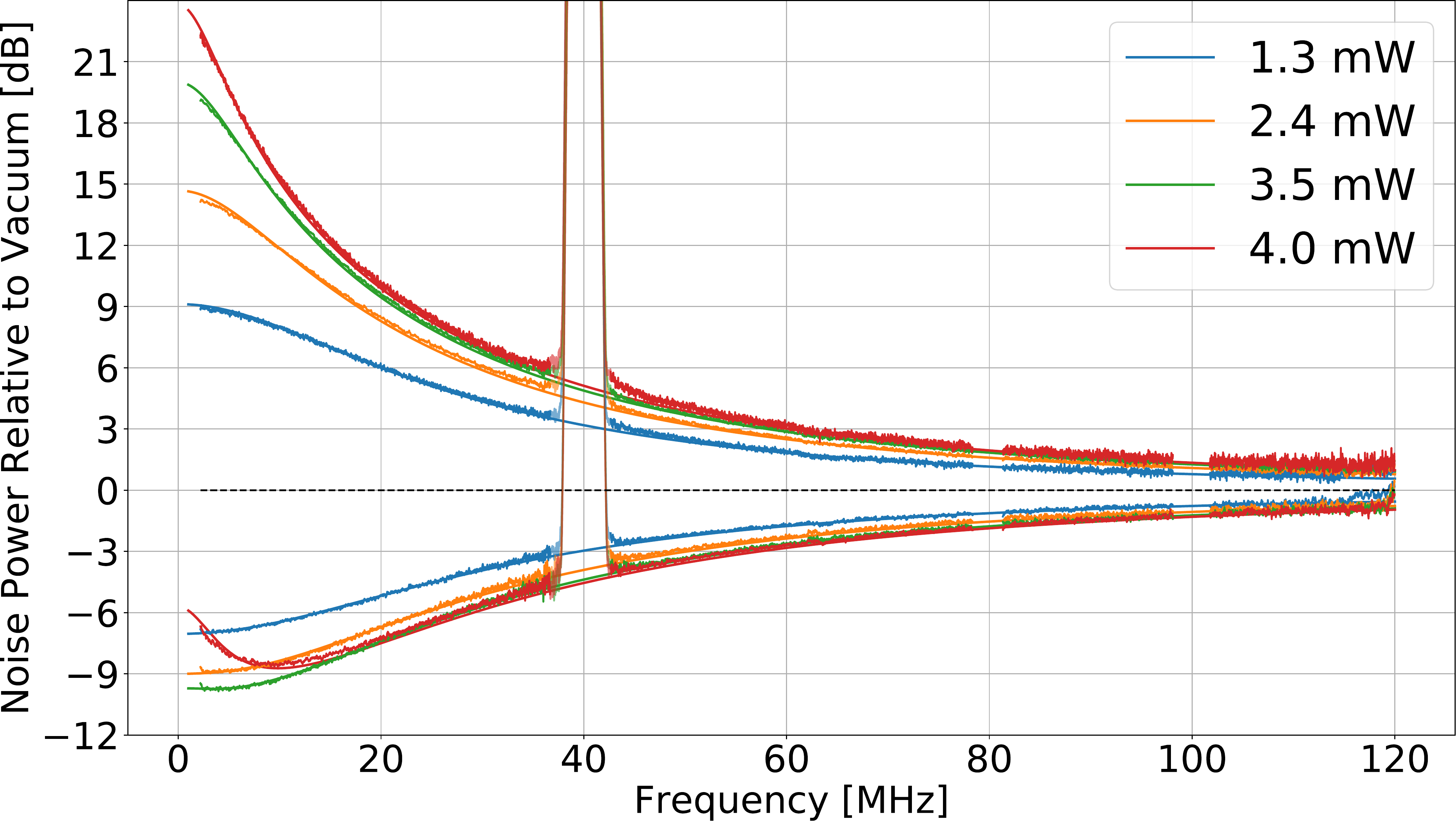}
    \caption{a) Squeezing and anti-squeezing relative to shot noise as a function of pump power at a side-band frequency of 5 MHz. The blue points are squeezing and the orange points are anti-squeezing. The theoretical model of Eq.~\ref{eq:squeezefull} is fitted to the data and plotted in solid lines. The purple dashed line is the fitted squeezing model in the absence of phase noise. The electronic noise, which is \SI{22}{\decibel} below shot noise, was subtracted from the data. The error-bars assume a 5\% error on the power. b) Spectrum of the squeezing and anti-squeezing from \SI[parse-numbers = false]{1-120}{\mega\hertz} for different pump powers. The traces are corrected for electronic noise and normalized to the shot noise. The thin solid lines are fits of Eq.~\ref{eq:squeezefull}. Bands around \SI{40}{\mega\hertz}, \SI{80}{\mega\hertz} and \SI{100}{\mega\hertz} are excluded from the fit as they contain the \SI{40}{\mega\hertz} up-shifted pilot tone and electronic pick-up of modulation signals, and the \SI{80}{\mega\hertz} and \SI{100}{\mega\hertz} peaks are removed from the figure.}
    \label{fig:sqz}
\end{figure*}

In order to characterize the squeezing performance, we first look at the \SI{5}{\mega\hertz} side-band frequency using a resolution-bandwidth of \SI{300}{\kilo\hertz} and video-bandwidth of \SI{300}{\hertz}. We measure for \SI{200}{\milli\second} and average each trace 100 times. The pump power is varied in the range \SIrange[range-units=single,range-phrase=--]{0}{4}{\milli\watt}, and the local oscillator phase is locked to squeezing and anti-squeezing, respectively. The measured values as a function of pump power are shown in Fig.~\ref{fig:sqz}a corrected for electronic noise, which is about \SI{22}{\decibel} below shot noise.

The squeezing increases with power until a maximum of around \SI{9.3}{\decibel} below shot noise at around \SI{2.5}{\milli\watt} pump power. After this, the squeezing begins to degrade with increasing pump power, suggesting that the measurement suffers from a significant amount of phase noise. The anti-squeezing is almost insensitive to phase noise and increases for all pump powers. The full power dependence of the (anti-)squeezing variance $\langle{\delta\hat{X}^2_{-}}\rangle$ ($\langle{\delta\hat{X}^2_{+}}\rangle$)  including phase noise can be modelled as (normalized to shot noise) ~\cite{Aoki2006}

\begin{widetext}
\begin{align}
\centering
\expval{\delta\hat{X}^2_{\pm}} \approx 1 + \eta_\text{esc}\eta_\text{opt}\mathcal{V}^2\eta_\text{QE}
\left(\pm\cos[2](\phi)\frac{4\sqrt{\frac{P_p}{P_p^\text{thr}}}}{\left(1\mp\sqrt{\frac{P_p}{P_p^\text{thr}}}\right)^2+4\left(\frac{\omega}{\kappa}\right)^2}\mp\sin[2](\phi)\frac{4\sqrt{\frac{P_p}{P_p^\text{thr}}}}{\left(1\pm\sqrt{\frac{P_p}{P_p^\text{thr}}}\right)^2+4\left(\frac{\omega}{\kappa}\right)^2}\right),
\label{eq:squeezefull}
\end{align}
\end{widetext}
where $\eta_\text{esc}\eta_\text{opt}\mathcal{V}^2\eta_\text{QE} = \eta_\text{tot}$ is the total efficiency, with $\eta_\text{esc}$ being the escape efficiency, $\eta_\text{opt}$ being the optical loss, $\mathcal{V}$ being the fringe visibility of the squeezing and the local oscillator, and $\eta_\text{QE}$ being the quantum efficiency of the photo diodes. $\omega = 2\pi\times \SI{5}{\mega\hertz}$ is the angular frequency of the measurement side-band and $\kappa = 2\pi\times \SI{66}{\mega\hertz}$ is the FWHM bandwidth of the OPO. Finally, $\phi$ is the RMS value of the phase noise. We note that the equation is only valid for small values of $\phi$. The solid lines in Fig.~\ref{fig:sqz}a are a fit to Eq.~\ref{eq:squeezefull}. From the fit, we extract a total efficiency $\eta_\text{tot} = \SI[separate-uncertainty]{0.92(1)}{}$. This is in pretty good agreement with our estimated efficiency $\eta_\text{tot} \approx  \SI{0.93}{}$ that comes from $\eta_\text{esc} \approx \SI{0.97}{}$ (estimated from coating specifications, ignoring scattering and absorption), $\mathcal{V} \approx \SI{0.99}{}$, $\eta_\text{QE} \approx \SI{0.99}{}$ and $\eta_\text{opt} \approx \SI[parse-numbers = false]{0.999^7}{}$. The optical loss comes from 5 mirrors and 2 lenses, all of which have an estimated efficiency of \SI{0.999}{}. The discrepancy is likely due to the actual quantum efficiency of the photo-diodes which is quite difficult to estimate precisely. Considering only the escape efficiency of the OPO allows us to determine the squeezing that is available directly out of the cavity, which can go as high as \SI{15}{\decibel} close to threshold.

From the fit, we also extract the RMS value of phase noise to be \SI[separate-uncertainty,multi-part-units = single]{19(1)}{\milli\radian}. This is quite high, and we believe this to be in part due to disturbances introduced by the fibers and in part due to back-reflections off the photo-diodes and back into the OPO. This issue requires further investigation as our setup seems to be limited by phase noise rather than loss. The theoretical model without phase noise in Fig.~\ref{fig:sqz}a indicates that our setup can reach detected squeezing levels beyond \SI{10}{\decibel} below shot noise.

Figure~\ref{fig:sqz}b shows the spectrum from \SIrange[range-units=single]{1}{120}{\mega\hertz} of the squeezing and anti-squeezing for varying pump powers including fits of  Eq.~\ref{eq:squeezefull}. 
As expected, the squeezing follows the bandwidth of the OPO, and is still present all the way out beyond \SI{100}{\mega\hertz}. Once again, for high pump powers the squeezing starts to degrade at low frequencies due to phase noise, and from the fits of the curves, we extract similar values for the efficiency and phase noise. A curious exception is the \SI{3.5}{\milli\watt} trace that surprisingly reaches all the way to \SI{9.6}{\decibel} below shot noise for frequencies of \SIrange[range-units=single,range-phrase=--]{1}{7}{\mega\hertz} and only contains about \SI{12}{\milli\radian} of phase noise from the fit.

\section{Summary}

In summary, we have demonstrated that a squeezed light source based on a traditional bulk squeezing cavity can be built with a small footprint on an area smaller than what is available in standard 19 inch boxes. We achieved this by replacing most free-space optics with fiber optics except for the in- and out-coupling of the OPO. The result was in particular made possible by a single-pass waveguide SHG module which provided enough pump power for the low threshold of only a couple of milliwatts, which in turn was due to the double-resonance of the OPO.

Further engineering on setup packaging will enable even smaller foot prints. A mobile turn-key squeezed laser in a transportable box might make squeezing a standard tool in quantum optics labs, which are ready to explore many yet undiscovered applications.

\section*{Acknowledgements}
The authors acknowledge support from the Innovation Fund Denmark through the Quantum Innovation Center, Qubiz as well as support from the Danish National Research Foundation, Center for Macroscopic Quantum States (bigQ, DNRF142). Furthermore the authors acknowledge the EU project CiViQ (grant agreement no.\ 820466) and the EU project UNIQORN (grant agreement no.\ 820474).

\bibliography{literature}

\end{document}